*Essay*

# The Meaning of “Big Bang”

**Emilio Elizalde**

Institute of Space Science, ICE/CSIC and IEEC, Campus Universitat Autònoma de Barcelona, C/Can Magrans, s/n, Bellaterra,
08193 Barcelona, Spain; elizalde@ice.csic.es

**Abstract**

What does “Big Bang” mean? What was the actual origin of these two words? There are many aspects hidden under this name, which are seldom explained. They are discussed here. To frame the analysis, help will be sought from the highly authoritative voices of two exceptional writers: William Shakespeare and Umberto Eco. Both have explored the tension existing between words and the realities they name. And this includes names given to outstanding theorems and spectacular discoveries, too. Stigler’s law of eponymy is recalled in this context. These points will be at the heart of the quest here, concerning the concept of “Big Bang”, which only a few people know what it means, actually. Fred Hoyle was the first to pronounce these words, in a BBC radio program, with a meaning that was later called inflation. But listeners were left with the image he was trying to destroy: the explosion of Lemaître’s primeval atom (an absolutely wrong concept). Hoyle’s Steady State will be carefully compared with inflation cosmology. They are quite different, and yet, in both cases, the possibility of creating matter/energy out of expanding space is rooted in the same fundamental principles: those of General Relativity. As is also, the possibility of having a universe with zero total energy, anticipated by R.C. Tolman, in 1934 already. It will be shown, how to obtain accelerated expansion from negative pressure; how to reconcile energy conservation with matter creation in an expanding universe; and a curious relation between de Sitter spacetime and Steady State cosmology. Concerning the naming issue, it will be remarked that, today, the same label “Big Bang” is used in very different contexts: (a) the Big Bang Singularity; (b) as the equivalent of cosmic inflation; (c) speaking of the Big Bang cosmological model; (d) to name a very popular TV program; and more.

## 1. Introduction

What’s in a name? What is hidden behind a particular name? What is its real content? In other words, what does a name contribute to what it names? We will conclude that, generically, nothing at all. A name is, in principle, merely a label. It does not change the nature or the quality of the thing it represents.

Such is the quintessence of the profound thought expressed by William Shakespeare in his world-famous play “Romeo and Julia”; when Juliet reflects on the fragrant rose, she is looking at (Figure 1):

*"That which we call a rose/By any other name would smell as sweet."*

**Figure 1.** (**a**) Juliet's phrase in William Shakespeare's play "Romeo and Julia". (**b**) Cover of the first edition of Umberto Eco's book "Il nome della rosa". Both images: fair use license.

In his more recent, and also famous title *"The Name of the Rose,"* Umberto Eco returns to the same paradox. The rose, symbol of love and beauty, has disappeared—only the name remains. Names remain, but their meaning fades. This is a central theme in Eco's novel: the meaning of a name often changes over time.

There is a remarkable conceptual kinship between both situations. Both Shakespeare and Eco explore the tension between words and the realities they name. The eventual conclusion is clear: names are, in general, just labels, simple stickers put there to identify things; even those given to great theorems or spectacular discoveries, as we are next going to argue.

This very important point will be at the heart of the investigation to be carried out here, around the concept of "Big Bang". We will see in detail how it will perfectly fit in.

Although, before we delve into this subject, it is necessary to add that there is a well-known principle—the so-called **Stigler's law of eponymy**—that goes even far beyond all the above, by stating that [1]:

***"No scientific discovery is named after its true discoverer."***

Stephen Stigler, a professor of statistics at the University of Chicago, stated it over forty years ago. Like every law, this one also has honorable exceptions; although its own author is not one of them. Stigler has always acknowledged that the actual discoverer of the law is not him, in fact, but the famous sociologist Robert Merton. However, as much as Stigler has repeated this truth a hundred times, he has not managed to get the law's name changed! It is a paradigmatic example that invites serious reflection.

Someone could immediately say that this law cannot be true, as such, that he or she knows cases in which it simply does not stand. This may be true, but it is a proven fact, when this is analyzed in depth and a sufficiently large data set is considered, that the number of cases that contradict the law are few, indeed; they are just the usual exceptions to the law. Let us list a couple of examples:

- ✓ The famous **Halley's Comet** was well known to astronomers since at least 240 BC.
- ✓ **Coulomb's law** was actually found by Henry Cavendish.
- ✓ The **l'Hôpital rule** in Mathematics is known to go back to Johann Bernoulli.
- ✓ The **Oort cloud** was first postulated and then described by another competent astronomer, Ernst Öpik.

- ✓ Öpik was also the first to calculate and publish the distance to Andromeda (**not Edwin Hubble!**), for which he obtained a value a factor of 2 better than Hubble's**.** Quite impressive. This fact should be stressed, because it is extremely difficult to find this explained in the popular literature.
- ✓ The all-famous **Hubble law** was first obtained and magnificently explained by Georges Lemaître in a publication in a Belgian journal, two years before Edwin Hubble obtained it. Only recently its name has been changed to the **Hubble-Lemaître law**. By the way, this is one of the very few examples of "reparation" that this author knows. It required a voting on-line of all members of the International Astronomical Union (IAU), in August 2018. In spite of this official name change, almost everybody continues to call it Hubble's law.
- ✓ The **Fermi Golden Rule** was first known and used by Paul Dirac, another great theoretical physicist.
- ✓ An additional and excellent example is the famous **Higgs boson**. To wit, most particle and high-energy physicists would agree that it should instead be called the "**ABEGHHK'tH**" **boson**, after the initials of Anderson, Brout, Englert, Guralnik, Hagen, Higgs, Kibble, and 't Hooft, at the very least. For, all of them (and probably a few more), contributed substantially to the formulation of this concept.

Anyhow, let us get back to the key issue. Where does a theory, a theorem, a discovery get its name from? It has been sometimes said that the term Big Bang began as an **insult** to a competing theory (this will be explained in Section 3). This might be true, at least to some extent; but that is only an irrelevant aspect of the main issue, as will be explained below. There is a lot more substance hidden here! And which, by contrast, is rather unknown. This is a paradigmatic situation: in too many popular accounts, even of very serious scientific issues, it is the anecdotic aspect, the irrelevant but spectacular detail what prevails; after being repeated one thousand times, as by an echo. And this also happens, regretfully, when the notice or information is completely false, or erroneous. Here we face one of the worst aspects of the information and AI era we are living in.

We will discuss all these issues in what follows, in relation with the Big Bang concept, starting from the very beginning. That is, from the point in which these two words, "Big Bang", were pronounced for the very first time in history. However, before doing that, and in order to immerse ourselves in the epoch and circumstances where this exceptional event happened, it is first necessary to recall an important piece of the history of modern cosmology. Namely, one corresponding to the previous years, which embraces what this author has termed as the first cosmological revolution of last Century. It goes from 1912 to 1932.

## 2. A Brief Account of the First Cosmological Revolution of the 20th Century (1912–1932)

The first revolution of modern cosmology, can be framed, to high precision, in the twenty-year period that goes from 1912 to 1932. That is, namely, from the most relevant astronomical discoveries of Leavitt and Slipher to those of Hubble, and includes the extraordinary theoretical advances made by Einstein, Friedmann, de Sitter, and Lemaître. It clearly came to its peak in 1929, with the publication of Hubble's results. There were, of course, many other contributions, but they cannot be fitted in this brief account [2]. Eventually, the scientific theory of the expansion of the universe having an origin was understood, step by step, and finally adopted by the main specialists in the celebrated Einstein-de Sitter model of 1932. Let us have now some (chronological) taste of what actually happened in that period.

This author has defended in different places that **the birth of modern cosmology occurred in 1912**. The reason is quite clear. For the first time in History, in that year the necessary tools become available to calculate: (a) the **distances**, and (b) the receding **speeds** of the celestial bodies at very large scale. Notice that here we are talking about cosmology, about the Universe as a whole, not about the astrophysics of nearby bodies.

We only need throw a look at the first map of the universe, of 1986, actually a slice of the cosmos capturing for the first time the third dimension, e.g., the distances to the far celestial bodies (Figure 2), to realize that, at the very large scales, celestial bodies appear as dots. Each of these points can contain billions of stars**.** The Universe is, thus, at this scale, a point distribution. The physical study of a point distribution starts by representing it in phase space. In other words, in order to proceed, the position and the speed of each point has to be determined, namely its six coordinates in phase space. When was it possible to obtain this information? Only in 1912, that is the reason why modern cosmology, that is, cosmology as a modern science, began that year. Not before, nor later.

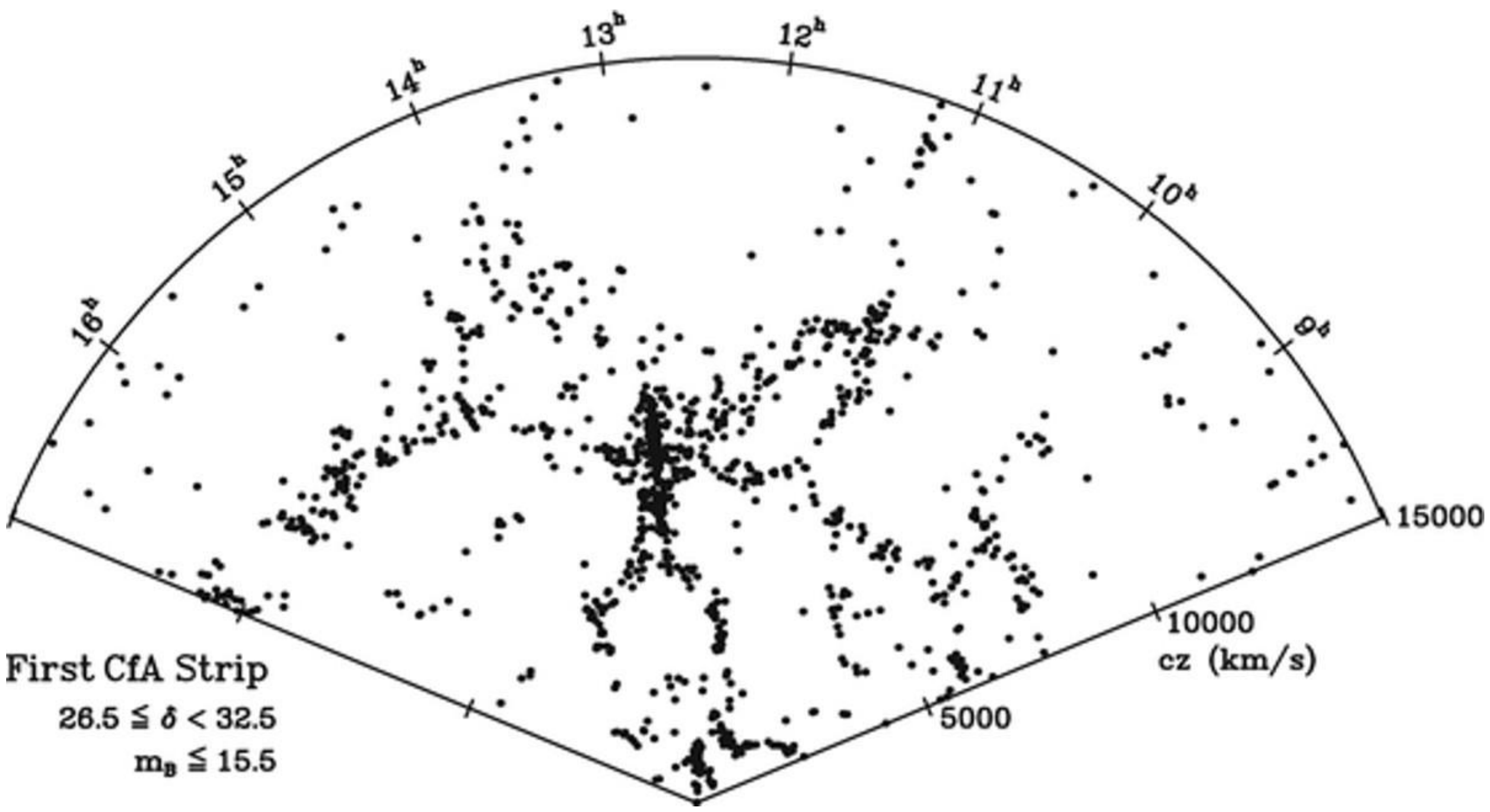

**Figure 2.** "A Slice of the Universe", March 1986, de Lapparent, V.; Geller, M. J.; Huchra, J. P.

Before this possibility was used, say namely a century ago, the Universe was considered to be: (a) **eternal** (it did not have an origin in the past); (b) **static** (or stationary, for very god reasons, since any physical system, under seemingly general conditions, will always evolve to a stationary state); and (c) **very small** (it reduced to the Milky Way, the thousands of distant nebulae that had already been spotted were all considered to be *inside our galaxy*). This absolutely wrong image of the cosmos was the result of not having been able, until then, to calculate the distances and speeds of the celestial bodies. Soon, this model of the Universe would change completely, in all of its conceptions. This is what the author of the present paper has baptized as **the first revolution in cosmology of the 20th century**, which did start in 1912. It coincided with the conversion of cosmology into a modern science. The main contributors to this revolution will be now listed.

*2.1. Henrietta Leavitt*

As it turns out, of the two tasks, the calculation of distances is by far the most difficult. Undoubtedly, it is the most difficult task in cosmology**.** Therefore, the merit of **Henrietta**

**Leavitt**, our first hero in this short account, is simply enormous. In 1912, after several years of collecting thousands of data, in particular from the Magellanic Clouds, which are satellite galaxies of the Milky Way, she discovered a relationship between the period of variability and the luminosity of Cepheid stars: the real luminosity of the star was proportional to the logarithm of the variability period of the luminosity of the star [3] (Figure 3).

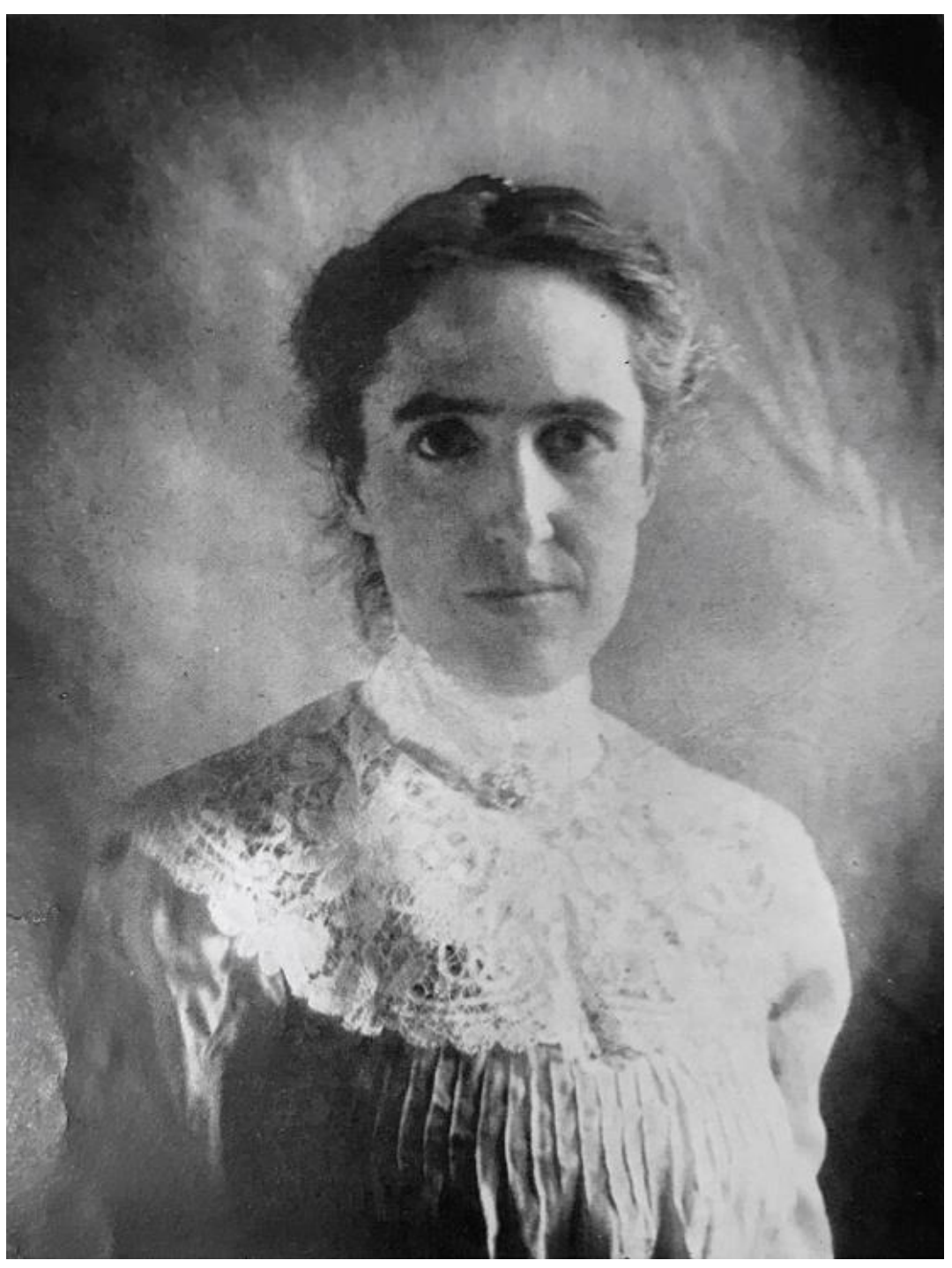

**Figure 3.** Henrietta Leavitt, age 30 (1868–1921). Public domain.

Leavitt's discovery was very simple, but it proved to be of enormous practical interest: **it is one of the most important in the whole history of astronomy** [4]. Leavitt's law made of the Cepheid stars the first **"standard candles"** in astronomy. Her result allowed astronomers to obtain the **distances to the faraway galaxies**, in the case when stellar parallax methods—ordinarily employed for close objects, were not useful, since the angles got incredibly small.

However, this was just a proportionality law, the overall scale had to be fixed. Fortunately, the astronomer Ejnar Hertzsprung calculated the distance to several Cepheids in the Milky Way, during the year after Leavitt reported her results. Having thus calibrated the scale, since then the determination of distances to far away Cepheids could be performed. Indeed, Cepheids were detected in other galaxies, what happened quite soon, as Hubble did in Andromeda (in 1922–1923), and obtained a result for the distance, which was ten times larger than any other distance calculated before within the Milky Way. A clear evidence that Andromeda, and by extension other spiral nebulae, too, were also galaxies, quite far from our Milky Way. Leavitt's discovery changed our image of the cosmos forever. The Universe became enormous, of a sudden.

But, what is the physical explanation of Leavitt's law? Why there is this relation between variability period and the intrinsic luminosity of such stars? Different mechanisms have been discovered during those more than hundred years, and now we know that, in fact, there are many different kinds of variable stars. Anyway, Cepheids continue, as of today, to be used to calculate distances. To oversimplify, let us just describe the valve effect, which occurs as follows. Helium has two possible ionization modes; partial ionization (when it loses one electron), and total (if it loses both). This happens to occur, as is easy to understand, in the upper and lower atmosphere layers, respectively, which are cooler and hotter owing to the distance to the star. But *He*+ is heavier and compresses the atmosphere of the star. This energy is mainly expended in doubly ionizing helium, in the lower level. But it turns out that *He*++ is opaquer and the star's radiation, unable to escape, pushes it to the upper layers, until it cools down again, becoming once more *He*+. This Helium being more transparent, it lets the star radiation through. And so, the star lights again: the luminosity peak of the Cepheid variable star occurs. When enough radiation has escaped and cannot keep the *He*+ layers on top, as *He*+ is heavier (as said) the upper layer falls down once more, to become *He*++ again, after a while. Another cycle starts, and so on. As explained, the mechanism is quite simple. Arthur Eddington produced one of its first versions. However, it has some problems, which will not be explained here for lack of space (they involve thermodynamic principles). Several other mechanisms for this pulsation have been found and such is still an interesting topic of study in astrophysics [5].

To summarize, Leavitt's method is extremely powerful for calculating distances. It constituted one of the main tools that were used by Hubble to obtain his law. And in the subsequent decades, by several generations of astronomers, with great success. Improved techniques along this line culminated in the supernovas SNIa as standard candles, which led to the discovery of the acceleration of the expansion of the cosmos (Physics Nobel Prize 2011). However, there we are entering the second cosmological revolution of the XX century, not to be considered here.

### *2.2. Vesto Slipher*

It was also in 1912, the year in which Leavitt published her findings, when the astronomer **Vesto Slipher** started a project, which would be no less transcendental (Figure 4). He obtained for the first time the radial speed of Andromeda, a nearby spiral nebula. He did this by using the optical Doppler Effect, accurately determining, to this end the deviations in the spectral lines (the so-called Doppler shifts), either towards the blue or towards the red. He was at the time at the Lowell Observatory in Arizona, and its 24-inch telescope was used to perform the observations. In some of the works of the present author [2,6,7], the enormous importance of Slipher's work has been duly stressed.

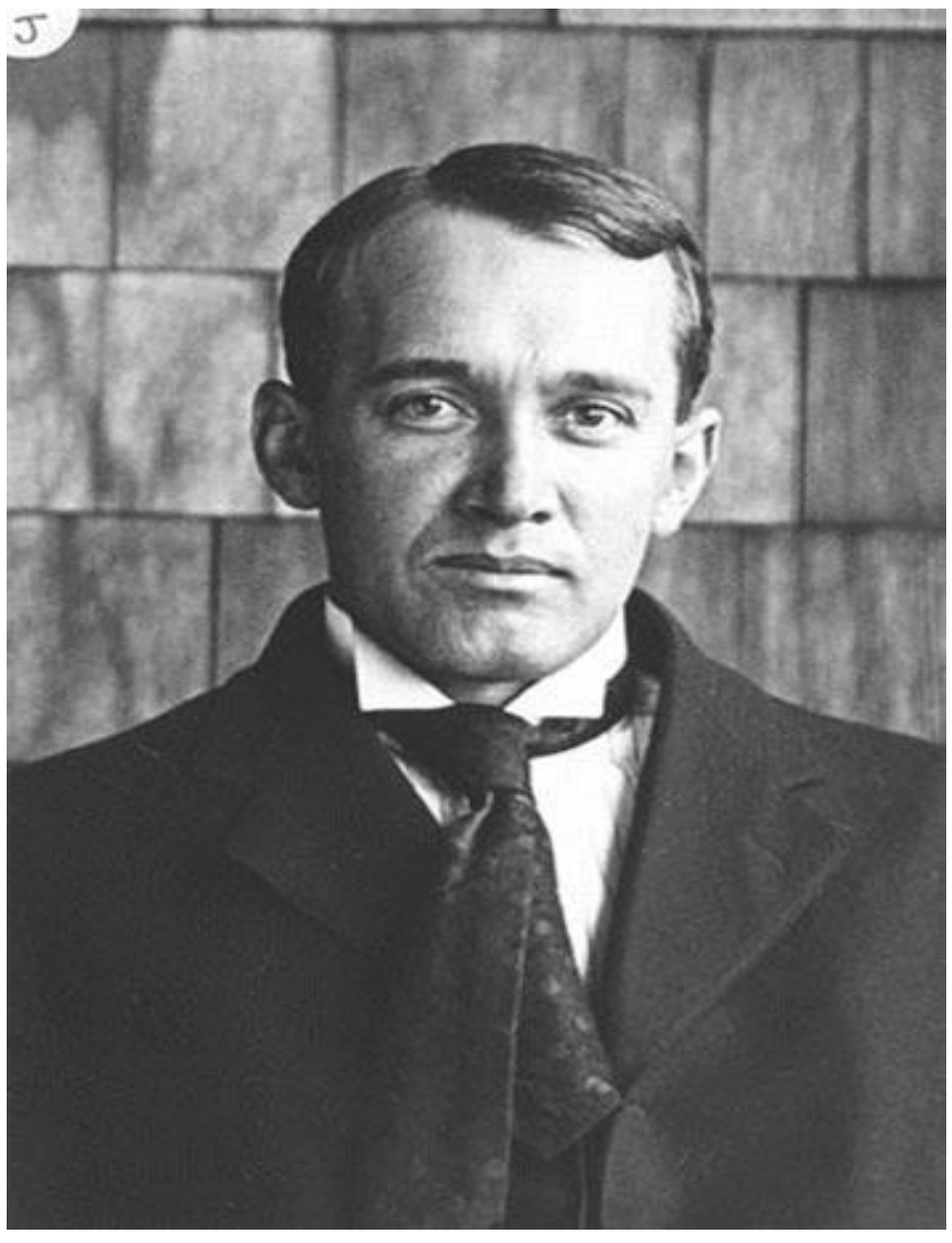

**Figure 4.** V.M. Slipher, astronomer at Lowell Observatory from 1901 to 1954. Unknown author, Lowell Observatory. Created: 1 January 1909. CC BY-SA 4.0.

It must be here emphasized that, in the earliest origins of the cosmological revolution the Lowell Observatory did play a key role. In spite that Hubble affirmed, throughout his life, that it was the work done at the Mount Wilson Observatory, alone, that really had transformed cosmology. This is not true, it does not stand a critical evaluation of the facts. Hubble, for one, forgot to admit—until the very last year of his life—that actually Vesto Slipher had opened his eyes to the revolutionary conclusion that the Universe could not be static! Hubble recognized, in particular, that *"the first step Slipher had taken was in fact the most important, making later progress in that direction, once already open, relatively simple."*

One can safely say that Hubble's character was Slipher's antithesis, in several important respects. Slipher was very shy; he did not publish many of his results, not even present them at scientific meetings. First, he had to be completely certain of what he had found. It then happened that many of his findings, now very much appreciated as important by specialists, remained unknown at that time, and even during many years after he died.

The observatory in Arizona was founded by the very rich astronomer Percival Lowell, who was also its first director. As such, he commissioned Slipher, as first duty, to assemble the spectrograph he had recently purchased for the observatory. With it, the spectral lines of light arriving from different kind of celestial objects could be measured. The optical Doppler Effect was already known since some decades ago. Luminous objects that move away from Earth, yield a spectrum that shifts toward the red part (a *redshift*), whereas objects approaching to us, produce a *blueshift*. The light arriving from luminous objects leaves traces from the different chemical elements the object contains, in the form of very characteristic spectral lines. Actually, one sees them as dark lines corresponding to certain wavelengths, which are typical of each chemical element. When the light source approaches the wavelengths move towards shorter ones (higher frequency, bluer ones), and when it moves away from the Earth, the spectral lines are displaced towards longer wavelengths (lower frequency, redder ones), respectively. This is completely analogous to the better known and commonly observed acoustic Doppler Effect, occurring when a noisy object (say a motorcycle) approaches first (acute sound) and then moves away from the observer (grave sound).

Slipher's project, aimed at obtaining the red and blueshifts of the spectra corresponding to spiral nebulae started in 1912, too, exactly the same year in which Leavitt published—actually with the signature of Edward Pickering—the final results she had obtained, after years of hard work. Slipher got, as a result of the spectral displacements, the radial speed of the celestial objects, approaching or moving away from the Earth (as Doppler displacements). Slipher had started his work in 1909, and by then he was an expert in handling the Lowell spectrograph. However, the light he got from the spiral nebulae under study was extremely dim. He had to increase the sensitivity dramatically. The size of the photographic plate had to be reduced to the area of a thumbnail. To the extreme that, in order to see the displacements, he had to use a microscope! In December 1912, Slipher arrived to a quite remarkable conclusion: Andromeda was approaching the Earth at the very high speed of 300 km/s! in accordance with the measured fact that the light coming from it was blueshifted. Such velocity was extraordinarily high, by then, about ten times the velocity obtained for other stars in the Milky Way! Slipher, who was always so rigorous, did not first trust his own result. Anyhow, Lowell kept encouraging him to continue and look at other spiral nebulae.

By August 1914, at the 17th meeting of the American Astronomical Society, which took place at Northwestern University (in Evanston, Illinois), Slipher gave a talk presenting the results from two years' work, corresponding to the velocities of 15 spiral nebulae. Only three of the analyzed nebulae were approaching to us, while all the rest were escaping at incredibly high speed. The average speed was 400 km/s. His presentation was very clear and professional and, according to the chronicles, it was received with a standing ovation [8]. This is not usual at a scientific conference, neither then not now, and that date has been since remembered as an important one in the history of astronomy. Among the numerous public attending there was a young astronomer, with the name Edwin Hubble, who got very impressed by those results, as he confessed towards the end of his life. Slipher was the first astronomer who could obtain the spectra of galaxies with a sufficiently good signal-to-noise ratio. He was the first who could measure the Doppler shifts reliably. And it is fair to say that his results shook the foundations of the model of the universe accepted by then: a static one.

Slipher continued his research and, by 1917, he had got data for the spectra of up to 25 spiral nebulae. From those, only three small nebulae and Andromeda (the closest objects, all of them) were approaching the Earth. The rest 21 objects, more distant, were moving away from us at high speeds. The conclusion Slipher extracted from these results was that our galaxy, too, should be moving in space at a very high speed and that, most likely, all these receding nebulae should be analogous to our own Milky Way. In short, other worlds like ours. And this was written by Slipher full eight years before Hubble detected the famous "Hubble Cepheid" in Andromeda, thus confirming the "island universe" conjecture of Immanuel Kant, Edgard Allana Poe and other thinkers.

As the same Edwin Hubble would finally admit (albeit only in the very last year of his life), Slipher was actually the first astronomer who pointed out, and very clearly, that something remarkable and quite strange happened in the Universe. How on Earth could it continue to be static and so small with all these distant objects that were escaping at such enormous velocities? Moreover, it has to be recalled that the table of redshifts produced by Slipher was one of the two ingredients used by Hubble to formulate his all famous speed vs distance law of 1929. The other ingredient, namely the table of distances to the nebulae, was actually due to Edwin Hubble, with a partial contribution from Milton Humason, who later calculated some more redshifts. Hubble did use them in 1931, in order to improve the earlier values. Although Milton Humason extended the spectral calculation to weaker galaxies, commissioned by Edwin Hubble, it is a clear fact that the astronomers at Mount Wilson could not have advanced as quickly without knowing already the pioneering results obtained by Slipher.

### 2.3. *Edwin Hubble*

**Edwin Hubble** was a very influential astronomer, probably, the most famous one of his generation (Figure 5). In the late 1920s, after a comparison of the redshift table for 25 spiral nebulae, which had been already published by Vesto Slipher in 1917 and was publicly available [9] (actually, it had already appeared in Eddington's famous book [10]), and his own table of distances to the same nebulae [11], he obtained the simple law, now very famous, that until very recently bore his only name, and was published it in 1929. In his article, he found a linear relationship between redshift and distances, which was interpreted by him, following Slipher's optical Dopplershifts, as velocities. He used all the data collected by Slipher, only adding a few more that were obtained by Milton Humason at Mount Wilson's observatory. Hubble found the distances to the nearest nebulae by using Cepheids as standard candles. For the objects that were more far away, he employed the brightest individual stars available, and assumed on passing that those were equally bright, for all nebulae. At even larger distances, he made use of the luminosities of the respective nebulae, as a whole object, as other astronomers did at the time.

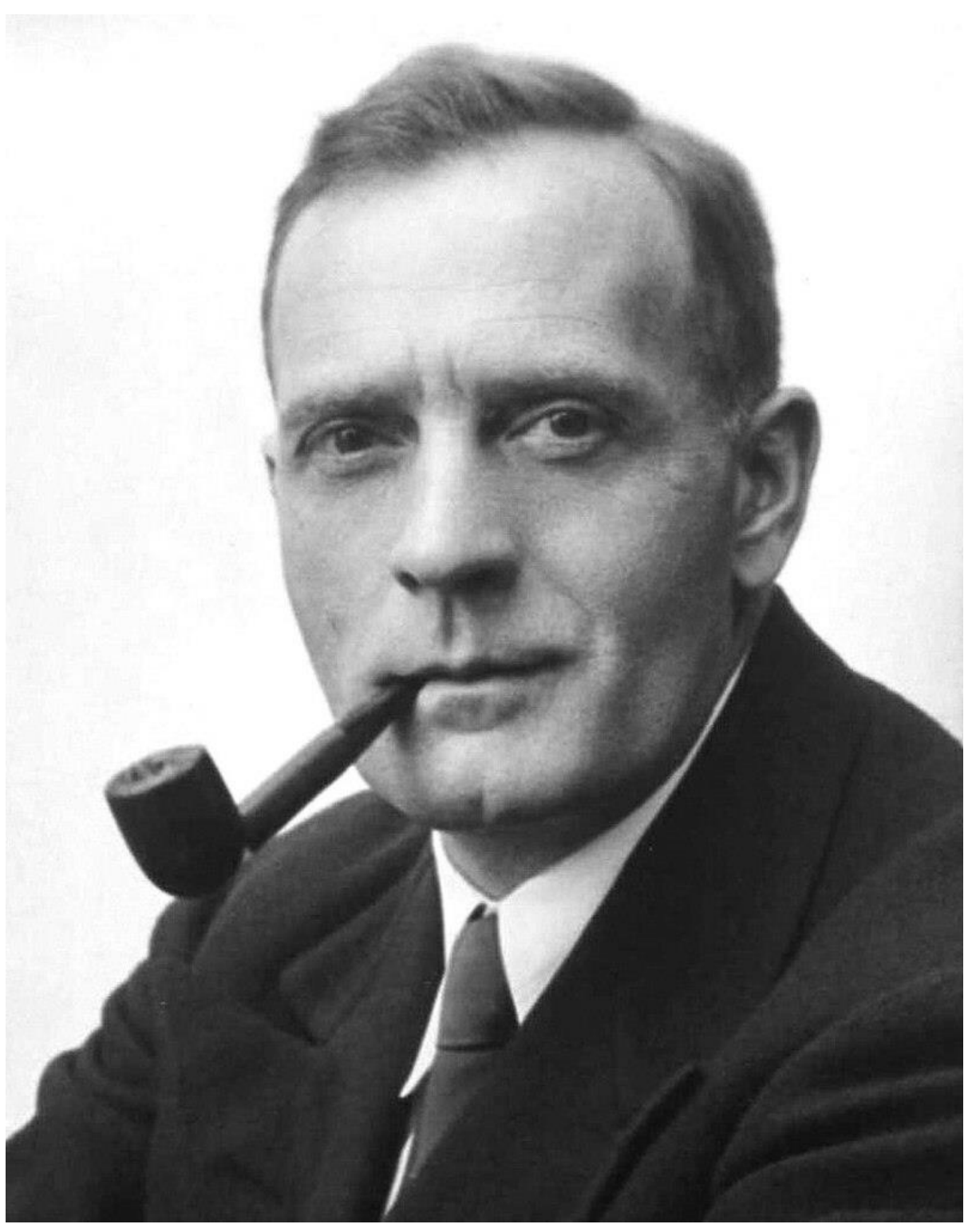

**Figure 5.** Studio Portrait of Edwin Powell Hubble. Photographer: Johan Hagemeyer, Camera Portraits Carmel. Photograph signed by photographer, dated 1931. Public domain.

When one places together the two tables, namely the one for the redshifts and the other one corresponding to distances, one suddenly realizes that the values fit smoothly into a straight line, $V = H_0 D$. This is sort of a children's play, no wonder that this was the proportionality that Lemaître obtained two years earlier, by just comparing the tables he graciously got from Slipher and Hubble. Here $H_0$ is just a proportionality constant, but a crucial one; probably the most important of all cosmological constants nowadays. Until quite recently, $H_0$ was called the Hubble constant. It is so, that Hubble wrote no mention in his paper, about the fact that Slipher was the sole author of the redshift table; indeed, his name does not appear anywhere in that seminal paper. This point is extremely important, since it explains why, for many years, and even today, in many references to Hubble's work, it is mistakenly considered that Hubble did produce alone all measurements of the two tables, namely the distances and the redshifts.

Hubble received, during his lifetime, many distinctions and recognitions. However, returning now to the name's issue—a key one in the present article—it has to be here mentioned that today, the name "Hubble" is best known to correspond not to his person, but to the space telescope launched by NASA in 1990 (and named after him). It has already celebrated 35 years in space. The very impacting images of the Hubble Space Telescope have reached all corners of the Earth and they have gone far into the depths of the human soul. In fact, it should be fair to say that they have done more for astronomy than many thousands of books, articles and presentations of the cosmos, worldwide.

### 2.4. *Georges Lemaître*

Our last hero in this necessarily short account of modern cosmology is none other than **Georges Lemaître**. He was mainly a mathematician, very good at solving differential equations (what he successfully did during his life), but not only that. He was also a real scientist, in spirit, who pursued for many years to construct a model of the cosmos in accordance with astronomical observations. No wonder that he became interested in paying a visit to the most important astronomers of his time. Since his stay at Cambridge University, UK, he got from Arthur Eddington, his host there, the commitment to build a valid cosmological model. All his efforts were focused, since then, on this goal. With this aim, he moved to the US, where he had conversations with Vesto Slipher, whom he visited at the Lowell Observatory in Arizona, and with Edwin Hubble, at Mt. Wilson, California (and also with other important scientists, as Robert Millikan, and others). Both Slipher and Hubble graciously shared with him their latest results and insights, they had got from direct astronomical observations. As already advanced, both handed to Lemaître their precious tables of speeds (as Doppler shifts) and distances of the spiral nebulae they had studied (most of the last, calculated using Leavitt's law).

The mathematical work Lemaître performed for his doctoral thesis at MIT culminated in obtaining a solution of the field equations of Einstein's GR. It describes a universe with mass and expanding uniformly, but which has neither origin nor end, due to the fact (not stressed by many authors) that there is an extra term, a logarithmic one, in Lemaître's solution. Ultimately, it was realized that this was just one of the solutions already found by **Alexander Friedmann** (Figure 6), in his famous paper of 1922 (it should be mentioned that Edward Kasner published in 1925 a related solution, which later became important, too). Although the solution found by Lemaître was not precisely the standard Friedmann solution (the one that Friedmann preferred), as many authors wrongly claim!

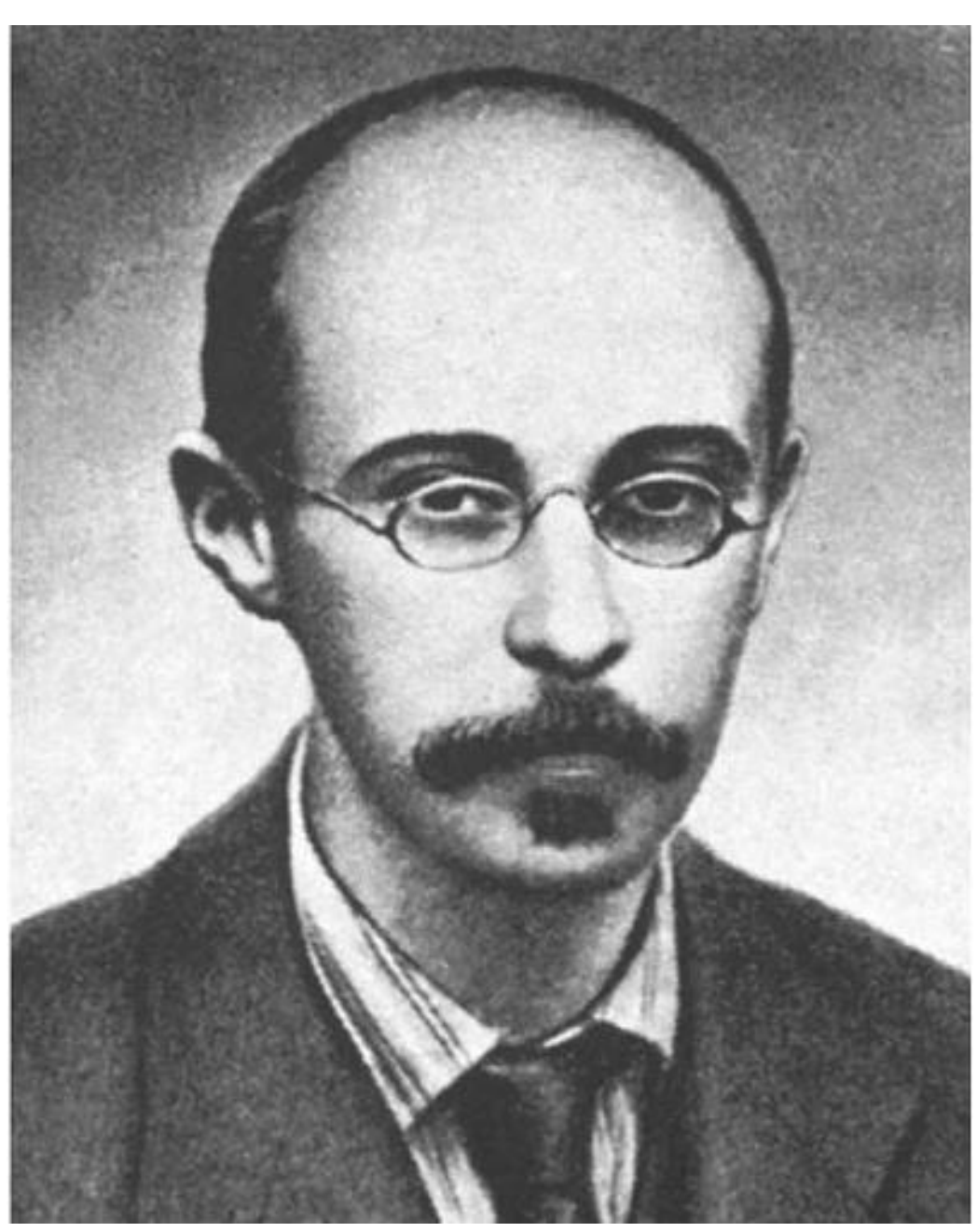

**Figure 6.** Russian mathematician and physicist Alexander Friedmann (1888–1925). Public domain.

After finishing his stay in Boston, and during two years more, Lemaître went on with his search for a model of the Universe fulfilling the observational results he had got from Hubble and Slipher. Finally, he was fully successful, in a very brilliant way, and published his conclusions in an article that was not then appreciated, but that would later earn him international reputation. He summarized in the paper all his work done at Harvard, MIT and Leuven. It was published in 1927, in an obscure Belgian journal, the Annales de la Société Scientifique de Bruxelles, under the title: "Un Univers homogène de masse constant et de rayon croissant rendant compte de la vitesse radiale des nébuleuses extra-galactiques" (A homogeneous universe of constant mass and increasing radius that explains the radial velocity of extragalactic nebulae) [12]. Lemaître presented in this work a revolutionary idea: the Universe was not static, but it was in fact expanding, according to the solution he got from Einstein's equations, and which agreed perfectly with the observational data of the astronomers Hubble and Slipher. To this end, Lemaître interpreted Slipher's table of spectral displacements as corresponding to Doppler shifts, and thus associated to receding velocities of the spiral nebulae. But he went farther than this, by saying that they were moving away not because of proper speeds but because of the Universe expansion. To this end, he matched the two tables, of redshifts and distances corresponding to 42 nebulae, and obtained a value for the Universe expansion rate (called Hubble constant now) quite close to the one Hubble got two years later, in 1929 (but only as an empirical constant). It is important to note, however, that Lemaître was not the first to state that the Universe might be expanding. That honor goes fully to Alexandr Friedmann, who definitely wrote about this possibility in his famous 1922 paper, which Lemaître apparently did not know about. In addition, Friedmann fiercely defended his dynamical solution in correspondence with Einstein, who stubbornly opposed the idea of an expanding universe for almost ten years.

In summary, Lemaître derived the same law as Hubble from the two tables he had got from him and Slipher. Conscious about that fact, he did never claim priority over his discovery. But, again, what is really remarkable is that, on top of this, Lemaître interpreted the redshifts in the correct way, as corresponding to velocities associated to the cosmic expansion (and not to displacements of the objects themselves). With his clear and precise intuition, he was at that point much ahead of the rest of astronomers and theoretical physicists of his time. This was amazing, just unbelievable in 1927. The data were at disposal of all astronomers, but none of them gave the correct interpretation of the same! Everything is to be found there, in his delicious article of only eleven pages, written in French, for anyone who wants to check it out. A warning: the official English translation—done a few years later, at the instance and with the help of Arthur Eddington—misses the section were Hubble's law is derived. After some intense debate, it now seems certain that it was Lemaître who did not consider this part to be important any more, given that the work by Hubble, with updated data, had been already published by then.

After accepting the Universe expansion, Lemaître had another brilliant idea: to look towards the past, back in time. An important issue to be considered here is that, in his 1927 model of the universe, time extended to minus infinity. It had no origin, owing to the presence of the logarithmic term in the solution he had found at MIT. Anyway, in 1931—when his latest work appeared in *Nature*—Eddington had already convinced him that such term was superfluous and led to incorrect conclusions. He showed him that the right solution was another one, previously obtained by Aleksandr Friedmann, too. Lemaître immediately changed his solution for precisely the one now known as "the Friedman solution." Shortly later, it was proven by the mathematicians Robertson and Walker that this one was the only possible solution to Einstein's field equations under the hypotheses of homogeneity and isotropy. And in the Friedman solution, the cosmos certainly had an origin in time!

Another comment is also in order. It was soon clear that the values of the expansion rate, both the one Hubble got, of 500 km/s/Mpc, and the value obtained by Lemaître, which was even worse, of 575, were too large. With them, the cosmos would had been only two billion years old, in the best of cases. But at that time, radioactive isotopes present in rocks already indicated that they were 4.5 billion years old, or more: the universe would have been then younger than the Earth! This was fully absurd, and was the main reason why Hubble, who was a very serious scientist, never believed in the expansion of the cosmos. However, this discrepancy does not seem to have posed any obstacle to Lemaître's imagination.

Such important fact is not pointed out very often, and it may seem contradictory: everybody admits Hubble discovered the Universe is expanding but it is a fact that he did not believe in what he had discovered! This is not so exceptional. Just recall that Columbus discovered America but he actually believed he was in Asia. It is a pity that Hubble did not live to understand, finally, that it was the calculated value of his constant, and not the model itself, what had to be seriously revised. Actually, Hubble's constant has had to be revised a huge number of times since then. We can even affirm that it is a value in permanent revision. Hubble's constant is, without any doubt, the most important of all cosmological constants and fixing it to good precision is an outstanding issue. Recently, tensions have been reported between its value at the very early Universe and in the present epoch. If confirmed, this could even lead to a dramatic change of the GR paradigm.

In 1931, Lemaître (Figure 7) was invited to London. While being there, and by using this time the "right" solution to Einstein's equations (the standard Friedmann's one), he put forth his brand-new proposal that the universe had been expanding all the time, starting from an initial stage in which it had had a very small size. He called this stage the "primeval atom" (it was very reminiscent of the "cosmic egg" conjecture appearing in several old cultures.) This time he published his theory in the prestigious journal *Nature* [13], and shortly later as an outreach article in *Popular Science* [14]. His argument was rather simple: as the Universe was expanding, by looking back towards the past, it must have been getting smaller and smaller, as it approached an initial time, from which it could not have contracted any further. All the energy and matter now existing in the universe had to be necessarily there, compressed in a very dense atom. Lemaître had some basic notions of the quantum theory of matter and he conjectured a big atom containing the whole universe. A "primeval atom" with a proportionally large nucleus, which disintegrated suddenly, at some point, in a huge explosion. And this outburst had led to the present distribution of energy and matter, now existing in the Universe, while the expansion would be continuing to this day. The cosmic rays, discovered by Victor Hess, back in 1912, were considered by Lemaître to be remnants of such an outburst.

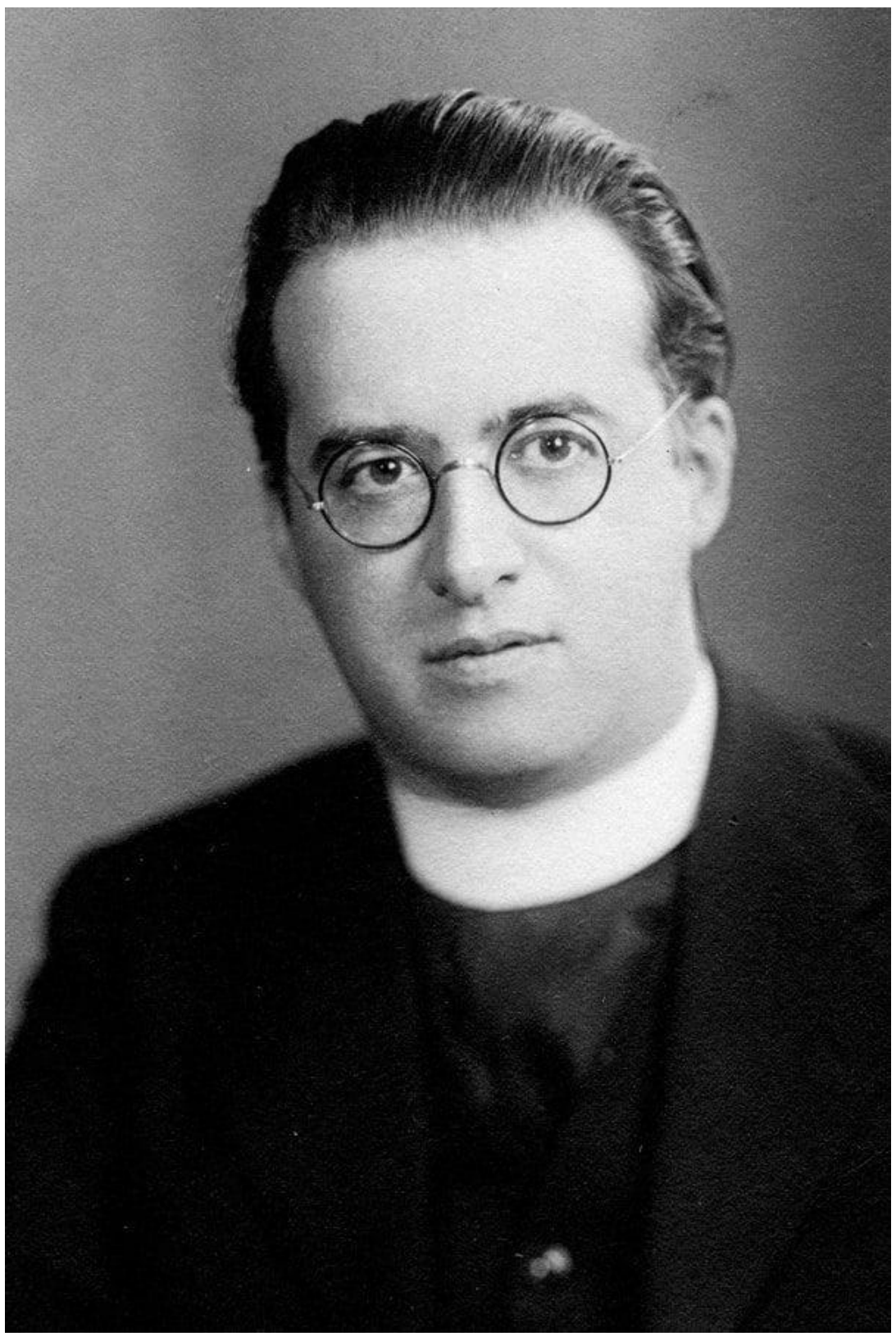

**Figure 7.** Picture of the years 1930 of Georges Lemaître (1894–1966). Public domain.

Lemaître's proposal was soon criticized by many of his scientific colleagues. Eddington, for one, found it *"very unpleasant,"* while Einstein's expressed the opinion that it was "unjustifiable from a physical point of view." This is in no way surprising, since Lemaître's knowledge of atomic and nuclear physics were very limited. Einstein had already changed, in this epoch, the good opinion he had maintained before about Lemaître, for some time. Previously, they had both been touring the USA giving conferences together, where Einstein had reportedly praised Lemaître's explanations of the evolution of the Universe as *"the most beautiful I have ever heard."* Anyway, no criticism against Lemaître's primeval atom was an obstacle for the popular acceptance of his fantastic model worldwide. Exactly the opposite happened! The simpler and more spectacular the better (even if it is completely wrong). Lemaître's model became extremely popular. It got engraved in the minds of common people, and it remains so to this day, almost hundred years later. It is just impossible to erase it from them.

The first colleague who definitely improved Lemaître's model and gave it some basic physical meaning, under the name of *"the Big Bang model,"* was George Gamow, a former student of Friedmann who had had to find a new thesis supervisor when Friedmann died prematurely in 1925. In accordance with the imaginative character that Gamow showed all his life, it was hardly surprising that he would choose the derogatory expression that Fred Hoyle had used when referring to this model, in an attempt to belittle it. We will explain this fundamental episode in all detail in the following sections. It is at the very heart of the present article.

## 3. And, in March 1949, Fred Hoyle Said: "Big Bang"

It was **Fred Hoyle**, a renowned English astrophysicist (Figure 8), who on 28 March 1949, first uttered the expression "Big Bang" [15]. He did so during an open lecture on cosmology, which he gave on the famous BBC's Third Program. Hoyle was one of the few scientists of the time who was invited to speak on this prestigious program while still being an assistant, and not yet a university professor.

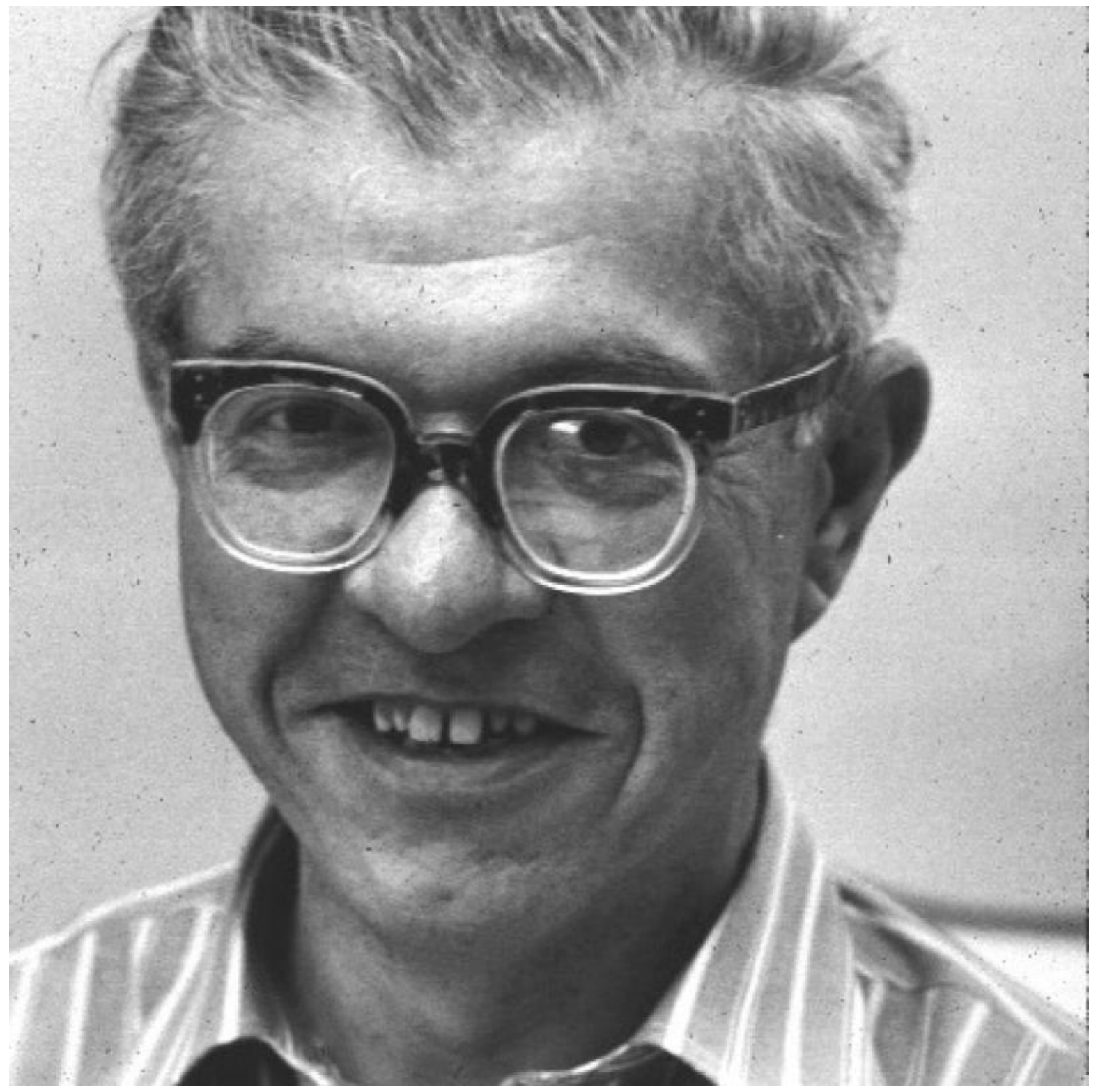

**Figure 8.** Fred Hoyle (1915–2001). Public domain.

Hoyle is well known today for his very important theory of stellar nucleosynthesis. He is the person who discovered that our bodies are all made of stardust.

At the start of the said BBC program, a broadcaster read the introduction:

> *"In this talk Fred Hoyle gives his reasons for thinking that matter is being created all the time, so that the universe must have had an infinite past and will have an infinite future."*

To what Hoyle added:

> *"I have reached the conclusion that the universe is in a state of continuous creation. … In a volume equal to a one-pint milk bottle about one atom is created in a thousand million years."*

Then he referred to the rival cosmological theories:

> *"We now come to the question of applying the observational tests to earlier theories. These theories were based on the hypothesis that all matter in the universe was created in one* ***big bang*** *at a particular time in the remote past."*

At some point, one could say that he tried to insult his colleagues, when he used the example of a climber:

> *"All routes taken by other theories attempting the unclimbed peak end in seemingly hopeless precipices."*
>
> *"The new way I am now going to discuss involves the hypothesis that matter is created continually."*

As for the method of creation, he invoked *"groundwork that has already been prepared by H. Weyl,"* (essentially, the General Theory of Relativity.)

**Important note.** We need to delve deeper into the era when things happened. The cosmological models that Hoyle considered as antagonists were: **the stationary model**, commonly accepted at the time (e.g., by Hoyle, Einstein, etc.)—in which the Universe had existed and had remained the same since always and forever—and **the new model of the expanding cosmos**—in which, looking back in time, the Universe had had an origin, as the Bible says. It is important to understand that in the stationary model the matter and energy of the cosmos have always existed—it was therefore not a duty to have to explain their origin—while, in the second case, as is clear from the book of Genesis, it is indeed necessary to explain the creation of matter/energy from nothing. This is the background of Hoyle's reasoning.

Hubble's discovery, immediately accepted by everyone, that distant nebulae are escaping at high speeds made it imperative that, in order to continue maintaining the stationary model, matter/energy had to be created constantly—although only in the very small proportion calculated by Hoyle—in order to always keep the density of the cosmos the same. This creation is possible, within GR, from an expansion (negative pressure) of the fabric of space (more specifically, in the reheating phase that occurs immediately after the expansion finishes). A possibility that, as an energy balance (in absence of a specific mechanism), was already clear since the publication, in 1934, of R.C. Tolman's book "Relativity, Thermodynamics and Cosmology". Here the possibility was already foreseen, too, that the total energy of the Universe might be zero (as a result of the cancellation of positive matter energy and negative gravitational binding energy), something that the latest observational data have demonstrated with good precision. These were very advanced concepts for their time and that inflationary theories made their own, as a starting point, half a century later. However, how and why this expansion of space occurred was another question, extremely difficult to answer.

Today we also know that, as early as 1931, Einstein himself had been working on a theory of this kind. He tried to create matter using the cosmological constant. His ultimate goal was exactly the same as Hoyle's: to keep the universe stationary, despite the recession of galaxies observed by Hubble in 1929. Einstein's manuscript was found (about ten years ago) sleeping peacefully in a drawer. The great genius was unable to come up with a convincing argument and never sent it for publication.

But Hoyle had gone much further. Together with Thomas Gold and Hermann Bondi they had designed an entire theory, that of the **Steady State Universe**—with an operator for the creation of matter—which for many years continued to compete head-to-head with the Big Bang model. Until the discovery, in 1964, of the cosmic microwave background radiation (CMB)—the first light in the Universe—which was a very important piece of evidence in favor of the Big Bang model. This model, it must be recalled, **still lacked the Big Bang** insultingly requested by Hoyle in his speech. It would not be incorporated, with all honors and under the name of cosmic inflation, until 1980, as we will see below in more detail. End of the note.

Taking up again Hoyle's exposition, he was fully convinced that the Big Bang he had proposed—as the only way to substantiate the model of Lemaître, Gamow, Alpher, Herman, and others—could never have happened, that it was absolutely impossible! And he expressed this thought clearly when, in a derogatory tone, he uttered for the first time in history the famous two words: "Big Bang."

The full text of his talk was published in April 1949 in the BBC weekly The Listener, and republished the following year, as part of a series of popular lectures entitled "The Nature of the Universe".

Anyone who wants to can easily access this information and relive that historical moment, in the precise terms in which this event occurred, word for word. But very few have taken the time to do so, and total ignorance, adorned with banal and absolutely wrong ideas spread like an oil stain, confusing the reality of the facts.

Here we will try, once and forever, to dot the i's and draw the pertinent conclusions in order to bring light to this important issue.

## 4. The Original Meaning of the Term "Big Bang"

Actually, the meaning of Hoyle's "Big Bang", was well established from the very first minute! Hoyle himself always defended this until the day of his death, occurred in August 2001: it was an impossible idea that was realized (by others) thirty years latter and is known today as "cosmic inflation".

Although it must be remarked that only a few—just those who were thoroughly familiar with the General Theory of Relativity (GR)—were able to understand him! That is, to understand, at the same time, both the theory that Hoyle defended of the continuous creation of matter in very small proportions, and also that of the creation, in a single stroke, of all the matter in the universe (or almost) in a very brief instant of time. Both matter creations are quite difficult to swallow, but the second does seem thoroughly impossible, as Hoyle rightly defended.

Matter/energy can, in principle, be created in GR, namely from an expansion, a dilation of the very fabric of space (an extraordinarily huge negative pressure that affects spatial coordinates). This is quite difficult to explain in just a few lines. In the theories of cosmic inflation it occurs in the so-called reheating phase, just after inflation ends; we will come to this important point, providing more details, later (see also [2,16]).

What eventually happened was that the overwhelming majority of radio listeners were left with the false image that Hoyle tried to destroy at any price. That is, the tremendous explosion of Lemaître's primeval atom or cosmic egg [17,18], which scattered all its matter and energy content throughout the rest of the Universe. This image is absolutely erroneous, it does not correspond at all to physical reality, but it is easier to 'understand' or 'digest'. In part, Hoyle should be blamed, since the name he chose, the expression Big

Bang, induces a lot of confusion, as Nobel Prize winner Jim Peebles never tires of lamenting.

And this brings us back to the very beginning of the topic. Let us recall the important conclusion we had already reached: **the name given to something must always be taken as a mere label**, with no direct meaning *per se*, without any relation (in principle) to the content of what it names. We should not try to deduce the content named from the name it bears! Thus, "Big Bang" does not mean anything, *per se*. It is only a label that was given, a long time ago, to an impossible idea that is presently a fully-fledged theory called "inflation" (Figure 9). Which is, in turn, another label (of course, the latter looks more appropriate for the phenomenon it names). This should be clear.

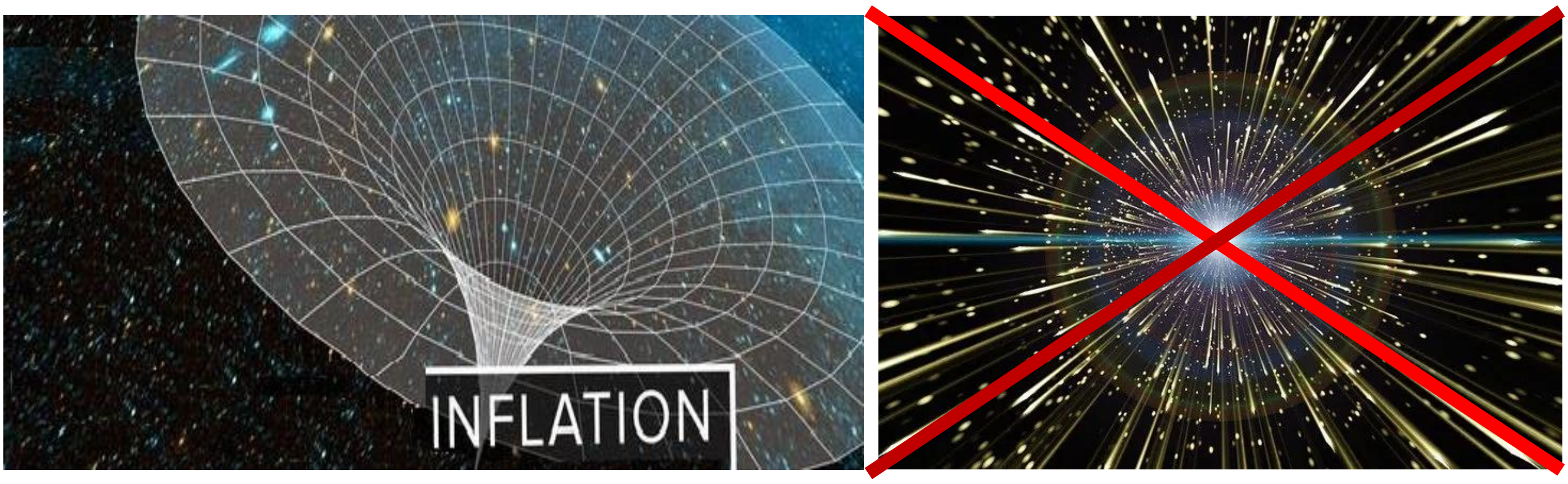

**Figure 9.** (**a**) Left figure. Inflation of a two-dimensional cosmos (1 time, 1 space dimensions). The time variable is the axis and the growth is exponential. (**b**) Right figure. Lemaître's explosion of the primordial atom or cosmic egg. This is a wrong concept (this is the reason why it is red crossed), it has nothing to do with what happened at the origin of our universe. Both images under fair use license.

Of a tip, let us look at an additional example. For some years now, we have known that the "Pythagorean theorem" was not actually first discovered by Pythagoras [19]. Lists of Pythagorean triples were already known and employed a thousand years before Pythagoras was born. And recently, an additional proof, a Babylonian clay tablet, known as YBC 7289 and dated to 1800 BC has been discovered. It is now kept at Yale University (USA) and apparently contains an actual demonstration of the theorem. Even then, the name has remained, no one has ever tried to change it. The point to be understood, once again, is that the name is just a label, in general. Pythagoras was certainly the first to prove the theorem in all its generality, he made extended use of it and made it popular in classical Greece. But the fact that we call the theorem by his name does not mean that Pythagoras was its first discoverer.

This is the important consideration that must be applied to our case: please, consider from now on the name "Big Bang" as just a label.

## 5. Original Big Bang Idea Was Realized by Cosmic Inflation

The prestigious writer John Gribbin, in the obituary he dedicated to Fred Hoyle—appeared in the newspaper The Independent on Friday, 17 June 2005, with the beautiful title: *"Stardust memories"*—stated that Hoyle would always be remembered as the main promoter of the Steady State Theory of the Universe, now completely discredited. But he also added this important reflection:

*"Everybody knows that the rival Big Bang theory won the battle of the cosmologies, but few (not even astronomers) appreciate that the mathematical formalism of the now-favored version of Big Bang, called inflation, is identical to Hoyle's version of the Steady State model."*

To say literally that the two mathematical formalisms are "identical" is actually not appropriate. In Steady State cosmology, the creation C-field maintains the matter/energy density constant, $\rho = \rho_0$, with $H = H_0$, whereas inflationary models employ a scalar field φ with a potential $V(\varphi)$ that drives exponential expansion $a(t) \sim e^{Ht}$ for a very short time.

Probably, what Gribbin had in mind (and Hoyle himself defended on several occasions) was that the driving idea behind the two theories was the same, that the same fundamental physical principle—as a possibility for the creation of matter and energy out of an expansion of space itself—is at play in both cases. As a matter of principle, this feasibility is inherent to Einstein's equations for GR, in a similar way that the possibility of conversion of matter into energy, and vice versa, was apparent in his world-famous equation for special relativity. In fact, the two formalisms (Steady State and inflation) are quite different, at the technical level, although they share the de Sitter metric as a formal framework, as will be explained in what follows.

### *5.1. Closer Comparison of Steady State and Inflation Cosmologies*

For clarification, let us compare now in a bit more detail Steady State and inflation cosmologies. The creation tensor of Steady State theory and the inflaton stress–energy tensor differ, in several important aspects. In short, the Steady State creation tensor is an *ad hoc* modification of Einstein's equations that violates local energy conservation to allow continuous matter creation,

$$G_{\mu\nu} = (8\pi G/c^4)\,(T_{\mu\nu} + C_{\mu\nu}), \quad (1)$$

where $C_{\mu\nu}$ is the creation tensor, not derived from an action principle. The creation tensor explicitly violates the usual local conservation law:

$$\nabla_\mu\, T_{\mu\nu} \neq 0, \quad (2)$$

a violation that is necessary to allow continuous matter creation.

On the contrary, the inflaton's stress–energy tensor arises from a standard dynamical field, with Lagrangian

$$L = \tfrac{1}{2}\,\partial_\mu\varphi\,\partial^\mu\varphi - V(\varphi), \quad (3)$$

and fully respects local energy–momentum conservation:

$$\nabla_\mu\, T_{\mu\nu\ \mathrm{total}} = 0. \quad (4)$$

Here, particle production occurs through energy transfer from the inflaton field during the reheating phase, right after inflation ends, not through a breakdown of conservation laws! The creation tensor does not correspond to any physical field or propagating degrees of freedom, while the inflaton is a dynamical field with physical degrees of freedom and quantum fluctuations.

To repeat, in Steady State, the creation tensor $C_{\mu\nu}$, introduced by Hoyle, Bondi, and Gold, is an extra tensor added by hand to Einstein's equations with the purpose to allow continuous creation of matter, in order to keep the density constant during expansion. In inflation cosmology, by contrast, the inflaton stress–energy tensor $T_{\mu\nu}(\varphi)$ comes from a physical scalar field φ with a given Lagrangian, and appears automatically from standard field theory in curved spacetime. While the creation tensor in Steady State violates standard local conservation, the inflaton field fully respects it. Particle creation occurs in this case via energy transfer from the field (reheating): energy flows between field and particles. This is a crucial distinction. Moreover, the creation tensor is phenomenological while the inflaton field is dynamical.

On top of this, Steady State cosmology is ruled out by observations such as the cosmic microwave background and primordial nucleosynthesis, while inflationary cosmology is strongly supported by observations of cosmic structure and microwave background anisotropies. To summarize, the Steady State theory was a failed attempt to realize the possibility, inherent to Einstein's GR, of trading space expansion for matter/energy. In contrast, inflation theories eventually succeeded, and in a brilliant manner, without breaking any basic physical law on the way.

### *5.2. Relation Between De Sitter Spacetime and Steady State Cosmology*

It is most interesting to realize how the de Sitter solution of Einstein's equations,

$$ds^2 = -dt^2 + e^{2Ht} (dx^2 + dy^2 + dz^2), \quad (5)$$

is related to Fred Hoyle's original Steady State formulation. The metric (5) represents de Sitter spacetime written in flat (k = 0) slicing. It is a solution of Einstein's equations with just a cosmological constant $\Lambda$ or, what is the same, to vacuum energy with an equation of state $p = -\varrho c^2$. As is well-known, key properties of this solution are: (i) exponential expansion, with $a(t) = e^{\wedge Ht}$; (ii) a constant Hubble parameter H; (iii) constant energy density; and (iv) spatial flatness.

On the other side, in their Steady State theory, Hoyle, Bondi, and Gold imposed the Perfect Cosmological Principle, requiring the universe to be homogeneous and isotropic in both space and time. This implies that $\varrho(t)$ = const., even though the universe is expanding. In such case, ordinary matter would dilute as $a^{-3}$, so Hoyle concluded that continuous matter creation had to occur. It was for this purpose that he introduced a creation tensor $C_{\mu\nu}$ and modified Einstein's equations in the form Equation (1). The creation tensor compensates for dilution, allowing constant matter density during exponential expansion.

As it turns out, an exponentially expanding, spatially flat universe with constant total energy density necessarily has the geometry of de Sitter spacetime. Thus, despite different physical interpretations, Hoyle's Steady State universe and the de Sitter solution **necessarily share the same spacetime geometry.**

Concerning the difference in interpretation, in modern cosmology expansion is driven by vacuum energy or a cosmological constant and no continuous matter creation is required. On the contrary, in Hoyle's interpretation expansion is driven by continuous matter creation, while the concept of vacuum energy was rejected as being unphysical. The metric itself is identical in both cases; only the source term differs.

It may look as an irony that Hoyle had actually rejected de Sitter spacetime as "empty" and unphysical (so did Einstein, for a while, too), yet his Steady State model corresponds mathematically to de Sitter spacetime written in matter-creating coordinates. This might be the correspondence Gribbin was referring to. However, as we see, it is only formal, not conceptual, anyway. Hoyle anticipated the geometry of inflation but rejected its physical interpretation, as stemming from vacuum energy.

The final word is that inflationary cosmology eventually showed that vacuum energy naturally produces a de Sitter phase, with matter creation occurring only temporarily, during reheating, rather than eternally. And this brilliant achievement resolved the conceptual problem without modifying Einstein's equations.

To summarize, Hoyle's Steady State universe is indeed geometrically equivalent to flat-sliced de Sitter spacetime with $a(t) = e^{Ht}$, but Hoyle was wrong in attributing the constant density to continuous matter creation, rather than to vacuum energy, a concept that he was unable to grasp (as also the importance of de Sitter spacetime).

*5.3. A Memorable Quote by Gamow*

Going now back on Gribbin's magnificent title, it turns out impossible not to mention the memorable quote that **George Gamow** (Figure 10)—an exceptional scientist with an excellent sense of humor—dedicated to Hoyle. In the chapter "New Genesis" of Gamow's autobiography "My World Line: An Informal Autobiography" (published posthumously in 1970) [20], he refers to Hoyle in these solemn terms:

> *"God was very much disappointed, and wanted first to contract the universe again, and to start all over from the beginning. But it would be much too simple. Thus, being almighty, God decided to correct His mistake in a most impossible way.*
>
> *And God said: "Let there be Hoyle." And there was Hoyle. And God looked at Hoyle … and told him to make heavy elements in any way he pleased. And Hoyle decided to make heavy elements in stars, and to spread them around by supernovae explosions."*

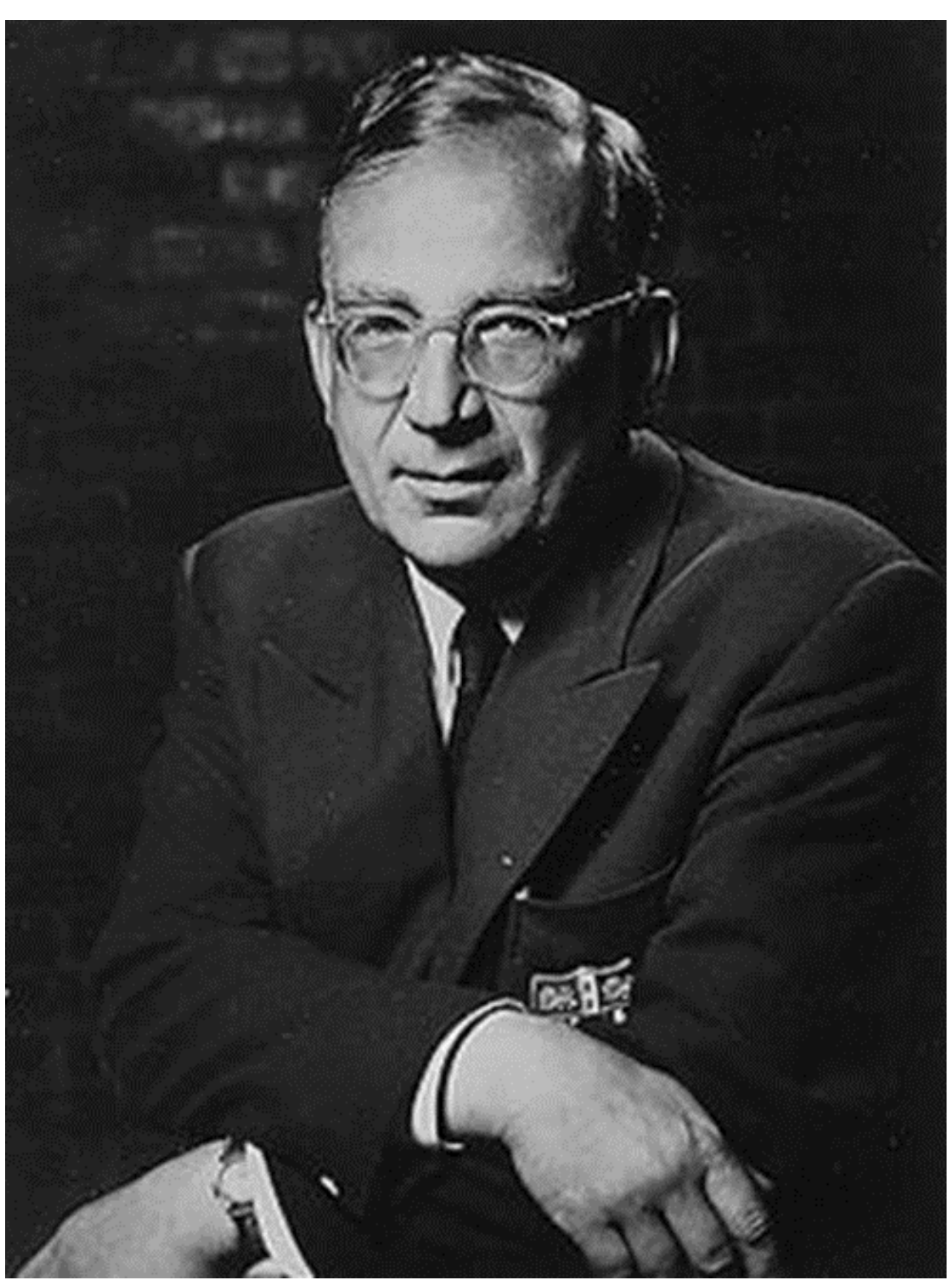

**Figure 10.** George Gamow (1904–1968) smoking a cigarette. Wikipedia Commons. Fair use.

And so, God was able to finish His work [21]. Notice that those ones are the real, spectacular, cosmic fireworks that Georges Lemaître, another of the main characters of our history, had referred to on so many occasions (although erroneously attributing them to the outburst of the primeval atom). We now know—thanks to the JWST space telescope—that supernovae explosions began to shape the evolution of the cosmos much earlier than it was thought until very recently. **Our three heroes connected by such incomparable universal beauty!**

One should here remark that, in the years to come, there were several crucial contributions from the rival teams in the subject of the Big Bang model. To begin with, Gamow, Alpher, Herman, et al. realized that the observed expansion of the Universe implied that, at a very early stage, it needed to have been hot enough in order that nuclear reactions could take place. Twenty years of important contributions on this issue finally led R.V. Wagoner, W.A. Fowler and F. Hoyle to the development, in 1967, of the **Big Bang**

**Nucleosynthesis** (BBN) theory [22], a very remarkable achievement. In fact, it still serves today as a most valued reference against which cosmological models have to be fitted. The final support to the model came, about the same time, from the fortuitous discovery of the Cosmic Microwave Background radiation (CMB). Anyway, in the period comprising the twenty years before that event, it had been clearly realized already that the temperatures of the BBN only allowed for the generation of a few lightest elements, up to Li at most. Therefore, it had become compulsory to explain: where did the rest of the many elements of the periodic table come from? And here it was Hoyle who led the field, with his spectacular theory of **Stellar Nucleosynthesis**, developed in his pioneering works of 1946 and 1954, and culminating in the landmark publication of 1957 by E.M. Burbidge, G.R. Burbidge, W.A. Fowler, and F. Hoyle, known as the $B^2FH$ paper [23].

## 6. On the Theory of Cosmic Inflation

For the benefit of the reader, a few paragraphs on the theory of cosmic inflation are now in order. As advanced, **the possibility** of matter/energy creation from a blow of the fabric of space (a negative space pressure) comes out most clearly from Einstein's General Theory of Relativity. This can be guessed, e.g., by looking carefully at the second Friedmann equation

$$\frac{\ddot{a}}{a} = -\frac{4\pi G}{3}\left(\rho + \frac{3p}{c^2}\right) + \frac{\Lambda c^2}{3} \quad (6)$$

On the left, we have the relative cosmic acceleration. That is, the second derivative of $a$, so-called cosmic scale factor, over $a$ itself, which is a typical cosmic length (say of 100 Mpc, or so) large enough in order to appreciate large-scale cosmological effects. Inside the brackets, $\rho$ is the matter density, the minus sign indicating that universe masses slow down its expansion (Newton's law). $\Lambda$ is the cosmological constant: if positive, it contributes to the cosmic acceleration. Also, in brackets, the pressure term, $p$, marks the crucial difference of GR with respect to classical mechanics. This pressure refers to the spacetime geometry, e.g., to the coordinate system itself, not to an ordinary pressure of a matter or energy component. It corresponds to compression (positive) or expansion (negative) of the very fabric of space (in the usual language employed in the popular literature). In a way, the reference system has an energy con tent. If $p$ is negative (expansion), one may say that we have a "negative gravity" component. This is the "miracle of physics N. 1" in the words of Alan Guth in his MIT lectures: in GR "gravity can be negative".

Going further, by looking in the interior of the round brackets, we realize that pressure goes together with matter/energy, this is the crucial point. The masses and energies in the universe can be increased (created indeed) at the expense of adding an equivalent amount of *negative* pressure, without changing the equation at all. In other words, according to the fundamental physical principles of GR, we can trade one for the other. In this way, the universe could have arisen from a state with zero net energy, and the present total energy of the universe could be exactly zero (as it seems to be actually the case!), if there would be as much matter as expansive pressure. These issues will be better explained below.

Such considerations are sort of a starting point in any standard lecture on inflationary theories. Unfortunately, we cannot go here into the technical details defining a particular theory, which constitute indeed the most important part of the whole issue. Nevertheless, let us just remark that, as a physical principle, as an in-principle possibility in total accordance with energy conservation, this issue was already clear in e.g., R.C. Tolman's book of 1934 and also, as is apparent from his speech, in F. Hoyle's argumentation when he said "Big Bang".

Anyway, we need make one thing clear. The fact that Tolman or Hoyle had such brilliant intuitions on the possibilities opened by GR physics does not at all mean that either of them should be considered as an actual pioneer of inflationary theories. For a good comparison, A. Einstein expressed very clearly in 1907 already that, as a consequence of the principles of special relativity, *"matter and energy are one and the same thing"*. That led him to write his most famous equation, $E = mc^2$, a few months later. But it would be unappropriated to consider him as the father of the atomic bomb, which is a technical consequence or practical realization of this very important principle. There was a long way and impressive amount of work by many clever people to fill the gap. We are talking here in similar terms, with respect to the actual development of cosmic inflation, by different specific ways, in comparison with the fundamental physical principles in which it is grounded.

For the benefit of the readers, and for completeness, we will now elaborate on these crucial points, from the modern perspective provided by our present knowledge of the subject.

### *6.1. How to Obtain Accelerated Expansion (A "Big Bang") from Negative Pressure*

To support the claim, sustained on Equation (6), that matter/energy can be created from negative pressure, we will here provide a derivation showing how the stress-energy tensor $T_{\mu\nu}$ with equation of state $p = w \varrho c^2$, where $w < -1/3$, leads to accelerated expansion.

For a homogeneous and isotropic universe, matter is modeled as a perfect fluid

$$T_{\mu\nu} = (\varrho c^2 + p) u_\mu u_\nu + p g_{\mu\nu}, \tag{7}$$

where $\varrho$ is the energy density and $p$ the pressure. Applying Einstein's equations to the Friedmann-Lemaître-Robertson-Walker (FLRW) metric yields the acceleration equation

$$\ddot{a}/a = -(4\pi G/3) (\varrho + 3p/c^2). \tag{8}$$

Substituting now $p = w \varrho c^2$, we get

$$\ddot{a}/a = -(4\pi G/3) \varrho (1 + 3w). \tag{9}$$

Here we see that acceleration requires $\ddot{a} > 0$. Since $G > 0$ and $\varrho > 0$:

$$\ddot{a} > 0 \Leftrightarrow 1 + 3w < 0 \Leftrightarrow w < -1/3. \tag{10}$$

In General Relativity, pressure gravitates, the effective gravitating density being (see Equation (6)): $\varrho + 3p/c^2$. We immediately see that negative pressure with $w < -1/3$ makes this quantity negative, producing repulsive gravity.

Imposing now energy conservation, $\nabla_\mu T^{\mu\nu} = 0$, we get that

$$\varrho \propto a^{-3(1+w)}, \tag{11}$$

and we conclude that, for $w < -1/3$, the energy density decreases slowly enough to drive accelerated expansion.

In summary:

$$p = w \varrho c^2 \text{ with } w < -1/3 \Rightarrow \ddot{a} > 0. \tag{12}$$

This explains why dark energy and vacuum energy cause cosmic acceleration.

### *6.2. How to Reconcile Energy Conservation with Matter Creation, in an Expanding Universe*

We will now see how, within General Relativity, energy conservation can be reconciled with matter creation in an expanding universe. We first address the issue of local vs global energy conservation.

In General Relativity, local conservation of energy and momentum is expressed as: $\nabla_\mu T^{\mu\nu} = 0$. This equation guarantees that energy is conserved in infinitesimal regions of spacetime. However, a globally conserved total energy exists only if spacetime has a global time-translation symmetry. But, an expanding FLRW universe does not have this symmetry, so global energy conservation is not well defined in this case.

Why does expansion break global energy conservation? While in flat spacetime, time-translation symmetry leads (via Noether's theorem) to conserved energy, in an expanding universe, the metric depends explicitly on time. Consequently, there is no global time-like Killing vector, and no invariant definition of total energy exists. Asking "where the energy comes from" is, therefore, not a well-posed question.

Let us have a look on the continuity equation and its thermodynamic interpretation. From $\nabla_\mu T^{\mu\nu} = 0$, in FLRW spacetime, it follows the continuity equation that describes how the energy density, $\varrho$, of the universe changes over time as it expands:

$$\varrho' + 3H\,(\varrho + p/c^2) = 0, \quad (13)$$

being $\varrho'$ the time derivative. This can be written as

$$d(\varrho\, a^3) = -(p/c^2)\, d(a^3). \quad (14)$$

Expansion works against pressure, so energy in a comoving volume is not conserved unless pressure vanishes.

Let us consider some basic examples. (i) For matter ($p = 0$), mass in a comoving volume is conserved. (ii) For radiation ($p = \varrho\, c^2/3$), energy decreases due to redshift. (iii) For vacuum energy ($p = -\varrho\, c^2$), the total energy increases as the universe expands. Anyway, no physical law is violated, because total energy is not defined globally.

For vacuum energy and the apparent energy creation, consider that $\varrho$ remains constant while volume increases, so total energy grows. But this does not imply a source of energy: gravitational energy is not localizable in GR, and no global conservation law applies.

In modern cosmology (ΛCDM, inflation), particle creation occurs naturally through reheating and quantum effects in curved spacetime, while respecting local conservation laws. By contrast, in steady-state cosmology, matter creation required modifying Einstein's equations, in a phenomenological way, without respecting conservation laws, as we have already explained before. Thus, General Relativity enforces local conservation of energy and momentum but does not require global energy conservation in an expanding universe.

In conclusion, matter creation in an expanding universe does not violate the energy conservation principle, because only local conservation laws exist, and an expanding spacetime does not admit a conserved total energy.

### *6.3. The Universe Can Have Zero Total Energy*

We here explain how the total energy of the universe can be exactly zero, within the framework of General Relativity and modern cosmology. As advanced above, total energy is a very subtle concept in GR. While energy is always locally conserved, global energy conservation only exists in spacetimes with time-translation symmetry. An expanding FLRW universe lacks this symmetry, so a globally conserved total energy is not uniquely defined. Nevertheless, in special cases—in particular, a spatially flat universe—it is indeed meaningful to discuss a total energy balance.

As positive energy contributions, we count all familiar forms of energy: (i) the rest-mass energy of matter; (ii) radiation energy; (iii) the vacuum (dark) energy. If only these contributions existed, the universe would have a large positive total energy.

As for negative energy contributions, in Newtonian physics gravitational binding energy is negative. And this is also true in General Relativity, although here gravitational energy cannot be localized as a density. More precisely, in a homogeneous universe, the mutual gravitational attraction of all matter produces a large negative contribution.

For a spatially flat universe (k = 0), we do obtain an exact cancellation. Indeed, in GR, the positive energy of matter, radiation, and vacuum energy is exactly balanced by the negative gravitational energy of cosmic expansion, namely

$$E_{total} = E_{matter} + E_{gravity} = 0. \quad (15)$$

The first Friedmann equation can be rearranged into a form resembling an energy balance between kinetic expansion and gravitational potential energy. For k = 0, the total energy term vanishes, corresponding to zero total energy. Inflation does generate enormous positive energy in matter and radiation during reheating, but it simultaneously produces negative gravitational energy, thus keeping an exact energy balance. Indeed, the cancellation remains exact, allowing the universe to grow without violating any conservation law.

This motivates the statement that **the universe could have arisen from a state with zero net energy**.

At this stage, the quantum cosmology viewpoint should be invoked. In quantum cosmology, the universe obeys the Hamiltonian constraint: H = 0. This implies that the total energy of the universe must vanish. Zero total energy is therefore a structural feature of General Relativity, not an added assumption, in this framework.

However, zero total energy does not imply global energy conservation or a classical creation from nothing. What it does imply is that positive and negative contributions cancel, allowing the universe to exist without net energy cost.

As a summary, the universe can have exactly zero total energy because the positive energy of matter and radiation is precisely balanced by the negative energy of gravity, in particular, in a spatially flat expanding universe. Anyway, **the role of curvature is important**, since in GR it is also a form of energy. Cosmological observations have confirmed, however, that its value is very small, compatible with zero, within an error of $10^{-5}$.

*6.4. Pioneers of Cosmic Inflation*

Back to the main point of the present section, when someone says "cosmic inflation" the first name that comes to mind is most probably **Guth**, maybe as second **Linde**, and probably as a third name **Starobinsky**. We will here take the opportunity to give due credit to everybody who contributed substantially to this theory (similarly to the aforementioned example of the Higg's boson). Perusing the literature on the subject, one easily encounters an interesting comment by John Peacock (written in 2017) where he pointed out that, in some sense, the first inflation paper was one by Erast Gliner, back in 1966 [24]. **Gliner** noted that an (unexplained) phase transition from a vacuum equation of state to radiation domination gave a scale factor that is exponential at early times, matching on to radiation domination. This is what was actually needed to solve the horizon problem, but this point was not duly emphasized in that paper.

In the real context of inflation as a theory, **Demosthenes Kazanas** (who sometimes appears as Dimitri in the literature) published in 1980 a paper in *The Astrophysical Journal Letters* proposing an early phase of rapid expansion [25]. The article addressed the horizon and monopole problems, what makes it clearly inflationary in spirit, even if the terminology and framework differ. His work was independent and contemporaneous with Guth's proposal (it was actually published a few weeks earlier).

As is widely known, **Alan Guth** introduced cosmologists to inflation at the 1980 Texas Symposium, although his paper was published in 1981 [26]. He introduced inflation as a solution designed to solve the horizon, flatness, and monopole problems. This is now called **old inflation**, because it needed to be improved, as soon serious problems were spotted.

Previous to all these papers, however, **Alexei Starobinsky** developed (in 1979–1980) an inflation-like scenario driven by **quantum corrections to gravity** [27,28]. Often regarded now as the earliest technically consistent inflationary model—though not framed in grand unified theories (GUT) language—it is sometimes considered among the most promising forms for an inflationary mechanism that could make a natural contact with fundamental theories of physics (to start with, because having to add quantum corrections seems unavoidable).

In 1981, **Katsuhiko Sato** published several papers proposing an inflationary universe driven by vacuum energy associated with phase transitions in GUT [29]. His work showed that supercooling during symmetry breaking could lead to a brief period of rapid exponential expansion. This addressed the horizon, flatness, and monopole problems, closely paralleling the motivations of Guth's original model.

Guth's old inflation model was substantially improved and made viable in papers appeared in 1982 and subsequently. In particular, **Andrei Linde** proposed the so-called **new inflation**, resolving the "graceful exit" problem of Guth's model [30]. And **Andreas Albrecht** and **Paul Steinhardt** (also in 1982) in joint work developed the new inflation theory too, independently from and nearly at the same time as Linde, explicitly emphasizing symmetry breaking and slow-roll dynamics [31].

To close the subject, the comments of some experts should be added, which point out that Guth might have been very lucky to be in the right place at the right time. They claim that Guth might have overheard Kazanas and his thesis advisor, David Schramm, discussing their brilliant idea that a de Sitter phase (of exponential expansion) in the early universe could have led naturally to an isotropic and homogeneous universe, in full agreement with cosmic observations. Guth might have benefited from this discussion between colleagues. And it is for certain that, at a couple of places in his own personal description of these days, Guth admits both that he was, in a sense, a newcomer to the field and that he overheard some discussions of colleagues. It is difficult to confirm how much these influenced his own discoveries. Nevertheless, what does seem strange is that, for decades, Guth never mentioned in his papers Kazanas's seminal contribution, published prior to his first work of 1981.

A possible corollary of all the above (and in relation with the naming issue) could be that, in parallel with the Higgs boson case considered at the beginning, cosmic inflation should go under the name (or label): **KGSSLAS**, or under some other (the reader's-preferred) permutation of this set of initials.

## 7. Discussion

1. **George Gamow**, a scientist (as explained) as brilliant as he was funny, mockingly **adopted the** ('insulting') **expression "Big Bang"**, to give a name to his theory (namely, that of Lemaître-Alpher-Gamow-Herman- …).
2. **Until 1980**, with the arrival of cosmic inflation, **the Big Bang theory** (originated in 1927, in Lemaître's now famous article in French, published in an 'obscure' Belgian journal, and later improved, as explained) **contained no Big Bang!** This was precisely the criticism that Hoyle had made about the model, expressing also that such a Big Bang was nevertheless impossible!
3. **D. Kazanas, A. Guth** (Figure 11)**, A. Starobinski, K. Sato, A. Linde, A. Albrecht, P. Steinhardt, … inserted the Big Bang (cosmic inflation)** in this Big Bang model.

4. This was **the Big Bang** (based on GR) **that F. Hoyle (quite rightly) said was missing from the theory!** Hoyle defended this point until the end of his life.
5. Although Hoyle's intuition was actually remarkable, both his conviction that the Big Bang was impossible, and his explicit way to address the matter creation issue proved to be wrong.
6. In contrast, the theory of cosmic inflation provided a brilliant solution to matter creation at the origin of the universe, in full accordance with all physical principles and fundamental conservation laws.
7. **Very important**. In addition to creating matter/energy, inflation (aka the real Big Bang), also **solved all the other serious problems** the model had: **homogeneity, causality, absence of monopoles**, etc.
8. Ultimately, it was the introduction of such true Big Bang, inflation, that completed and gave physical meaning to the Big Bang model. Full stop.
9. The name **Big Bang** is very **misleading** (as Jim Peebles often remarks). Anyhow, like all names, it should be understood as a simple label, nothing more. And this fixes the issue!
10. The "Bang" is actually an expansion, a brief "puff" that widened the fabric of space, affecting the entire Universe, for a very short time (not an explosion in a specific place).
11. It was certainly "Big", enormous, according to the number of e-folds (50 to 60) that the Universe expanded in an infinitesimal time.
12. In short, the Big Bang model got its name from a cynical criticism by Hoyle, who said (with all the reason in the world) that such model was missing a stage, as essential as it was impossible: a Big Bang! When this was incorporated, a model ***"comme il faut"*** was finally created [32–34].
13. The most universally widespread meaning of "Big Bang," **today**, is as the **"Big Bang singularity."** This is the result of important work by **Roger Penrose**, **Stephen Hawking** and others, who demonstrated that GR inevitably leads to a mathematical singularity at the origin of the cosmos, known now as the "Big Bang singularity" [35–41].
14. Nevertheless, this is by no way the end of the story. As Hawking himself acknowledged on his personal website (while he was alive), **GR is not valid at the origin of the Universe**. It is no longer valid much earlier, at subatomic scales! Such singularity is only mathematical, it has no physical meaning. But encodes important information: it tells us clearly that **we badly need new physics** [42]—which we do not yet possess—**beyond GR and quantum field theories.**

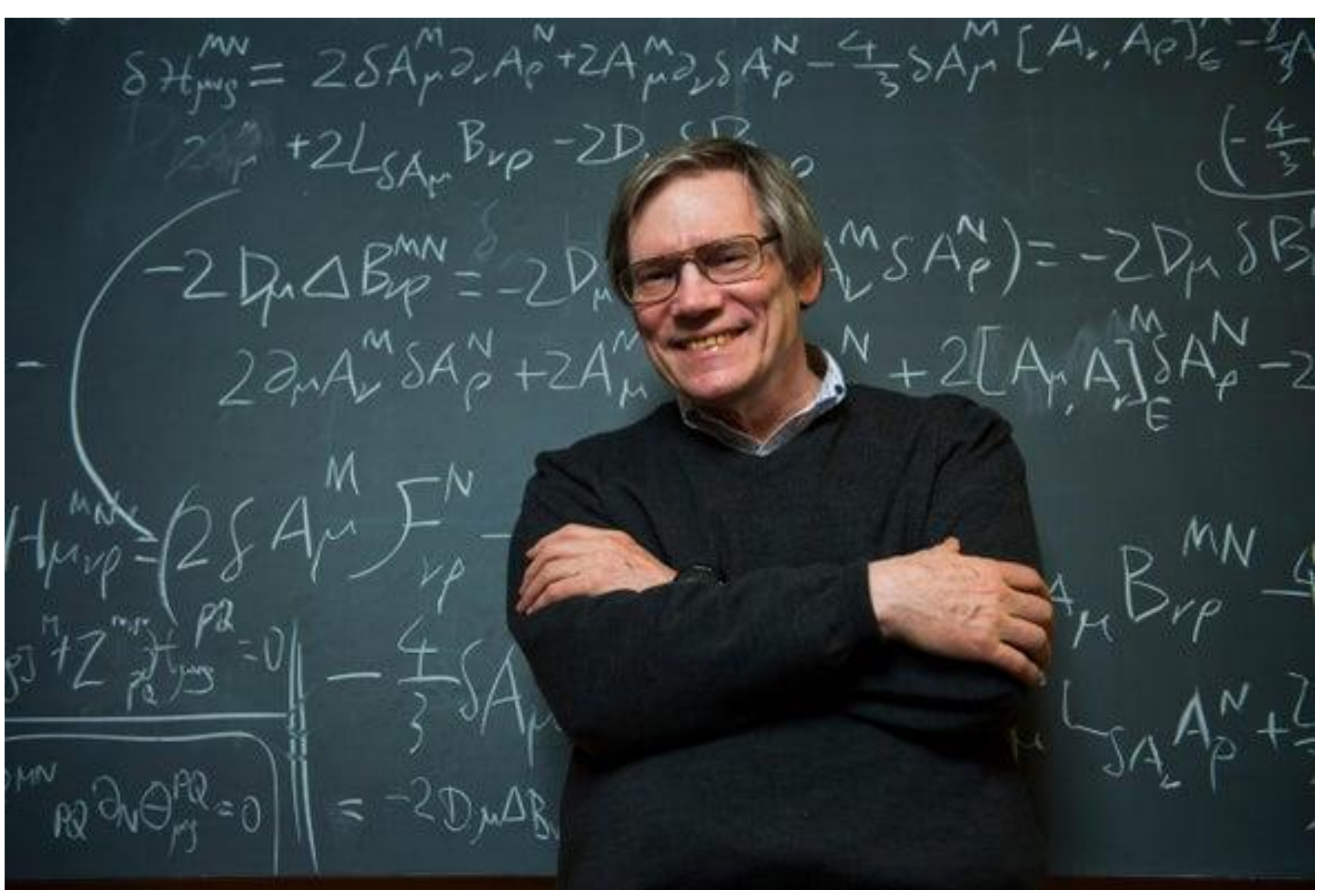

**Figure 11.** Alan Guth (1947–). Credit: Deanne Fitzmaurice/National Geographic Image Collection/Alamy.

## 8. Conclusions

I. **For over 50 years, the Big Bang model did not include any Big Bang**. As a consequence, **many problems accumulated**—besides the fact, to begin with, that **it was not complete**, since it was missing the initial stage of matter/energy creation. Actually, **it should have been called *the no-Big-Bang model!***

II. In 1981, the true Big Bang (cosmic inflation) was incorporated into the Big Bang model. Like by magic (although with a very clear purpose, actually), this solved all the problems at once!

III. The last stage, still pending, will be to rigorously demonstrate that the Big Bang did in fact occur!

IV. A more appropriate term (in the author's opinion) would be probably **Big Blow**: a powerful blow that suddenly expands the very fabric of space (say, the 3-dimensional surface of a **4-dimensional** sphere or ellipsoid, **cosmic space-time**).

V. **Important warning**: when people talk now about "Big Bang," and that's it, most likely they are referring to the **"Big Bang singularity."** A mathematical singularity without any direct physical meaning.

VI. Today the label "Big Bang" is used in several **different contexts**: (a) Big Bang **Singularity**; (b) as the equivalent of **cosmic inflation**; (c) speaking of the Big Bang **cosmological model**; (d) to name a popular **TV program**; ...

**Funding:** This research received no external funding.

**Data Availability Statement:** No new data were created in this work.

**Acknowledgments:** The author thanks Jose Gaite and Beatriz Gato for useful discussions. The demanding comments of the editor and of four anonymous referees have decisively contributed to the improvement of the manuscript. This work has been supported by the program Unidad de Excelencia María de Maeztu CEX2020-001058-M, and by the Catalan Government, AGAUR project 2021-SGR-00171.

**Conflicts of Interest:** The author declares no conflict of interest.